\documentclass[a4paper,11pt]{article}
\pdfoutput=1 %
\usepackage{jcappub} %

\usepackage[T1]{fontenc} 

\usepackage{graphicx}
\usepackage{hyperref}
\usepackage{soul}
\usepackage{amsmath}
\usepackage{lineno}
\usepackage{multirow}
\usepackage{color}
\usepackage{MnSymbol}

\title{Gamma-ray emission from the Seyfert galaxy NGC~4151: Multi-messenger implications for ultra-fast outflows}

\author[a,b,1]{E. Peretti,\note{Corresponding author.}}
\author[b,c,2]{G. Peron,\note{Co-corresponding author.}}
\author[d,e,f,g,h]{F. Tombesi,}
\author[e]{A. Lamastra,}
\author[e,i]{F.~G. Saturni,}
\author[b]{M. Cerruti,}
\author[a]{M. Ahlers}

\affiliation[a]{Niels Bohr International Academy, Niels Bohr Institute, University of Copenhagen, Blegdamsvej 17, 2100 Copenhagen, Denmark}
\affiliation[b]{Université Paris Cité, CNRS, Astroparticule et Cosmologie, 10 Rue Alice Domon et Léonie Duquet, 75013 Paris, France}
\affiliation[c]{INAF - Astrophysical Observatory of Arcetri, Largo E. Fermi 5, 50125 Florence, Italy}
\affiliation[d]{Physics Department, Tor Vergata University of Rome, Via della Ricerca Scientifica 1, 00133 Rome, Italy}
\affiliation[e]{INAF -- Astronomical Observatory of Rome, Via Frascati 33, 00078 Monte Porzio Catone (RM), Italy}
\affiliation[f]{INFN -- Rome Tor Vergata, Via della Ricerca Scientifica 1, 00133 Rome, Italy}
\affiliation[g]{Department of Astronomy, University of Maryland, College Park, MD 20742, USA }
\affiliation[h]{NASA Goddard Space Flight Center, Code 662, Greenbelt, MD 20771, USA}
\affiliation[i]{ASI -- Space Science Data Center, Via del Politecnico snc, 00133 Rome, Italy}

\emailAdd{enrico.peretti.science@gmail.com}
\emailAdd{giada.peron@inaf.it}
\emailAdd{francesco.tombesi@inaf.it }
\emailAdd{alessandra.lamastra@inaf.it}
\emailAdd{francesco.saturni@inaf.it}
\emailAdd{cerruti@apc.in2p3.fr}
\emailAdd{markus.ahlers@nbi.ku.dk}

\abstract{The nuclear activity typical of Seyfert galaxies can drive powerful winds where high-energy phenomena occur. In spite of their high power content, the number of such non-jetted active galactic nuclei (AGN) detected in gamma rays is very limited.
4FGL~J1210.3+ 3928, a source recently discovered by the Fermi-LAT telescope, is spatially consistent with the blazar 1E~1217.9+3945 and NGC~4151, a Seyfert galaxy located at about 15.8 Mpc known for hosting ultra-fast outflows (UFOs) in its innermost core. 
We show that the localization of 4FGL J1210.3+3928
might be affected by fluctuations due to a superposition of the two nearby sources. 

We explore the possibility of NGC~4151 to be a high-energy source and
we conclude that particle acceleration at the UFO wind termination shock can explain the luminosity and spectral shape of the observed gamma-ray flux, whereas the multiwavelength spectral energy distribution of 1E~1217.9+3945 disfavors it as the dominant GeV gamma-ray counterpart.
Interestingly, NGC~4151 is also spatially coincident with a weak excess of neutrino events identified by the IceCube neutrino observatory. We compute the contribution of the UFO to such a neutrino excess and we discuss other possible emission regions such as the AGN nearest neighborhood.}

\begin{document}
\maketitle
\flushbottom

\section{Introduction}
\label{sec:intro}

Ultra-fast outflows (UFOs) have been discovered through observations of highly blue-shifted Fe XXV and Fe XXVI K-shell absorption lines in the X-ray spectra of active galactic nuclei (AGN) and quasars~\cite{Chartas_2002,Chartas_2003,Chartas_2009,Pounds_2203,Dadina2005,Markovitz2006,Braito2007,Cappi2009,Reeves2009,Giustini2011,Gofford2011,Lobban2011,Dauser2012}. In particular, the definition of UFOs was first adopted in order to uniquely identify AGN-driven outflows featuring high ionization level and blue-shifted lines with associated velocity $\gtrsim 10^4\,{\rm km}\,{\rm s}^{-1}$~\cite{Tombesi_2010}. 
A systematic search performed in \cite{Tombesi_2010} showed that UFOs are common in nearby ($z\lesssim 0.1$) Seyfert galaxies  with a detection fraction $\gtrsim 40\%$. Moreover, a photo-ionization modeling performed in \cite{Tombesi_2011} allowed the authors to constrain the typical UFO velocity in the range $u \sim 0.03c-0.3c$. In addition, the authors of \cite{Tombesi2012} estimated that the typical distance of UFOs from the central supermassive black hole (SMBH) is of the order of sub-parsec scales and, in the same work, the typical values for mass loss rate ($\Dot{M}\sim 0.01-10\,{\rm M}_{\odot}\,{\rm yr}^{-1}$) and kinetic power ($\Dot{E}_{\rm kin} = \Dot{M} u^2/2 \sim 10^{41}-10^{46}\,{\rm erg}\,{\rm s}^{-1}$) were also derived.

UFOs could originate from accretion disc winds launched close to the central SMBH in AGN~\citep[][]{Tombesi_2010, tombesi2013, tombesi2015, Gofford2013, Laurenti2021}. 
They are mildly relativistic flows characterized by a wide opening angle \citep[][]{Nardini2015}, typically observed in the nearest parsecs from the SMBH \cite{Laha_2021}. Similar to other astrophysical diverging flows, UFOs can feature a bubble structure~\citep[][]{Faucher-Giguere-2012} characterized by an inner wind termination shock and an outer forward shock~\citep[][]{Zubovas2012, Costa2014}.
Recently, the authors of \cite{UFO_1} (hereafter referred to as P23) studied the possibility of accelerating particles at the wind termination shock of UFOs. In fact, ideal conditions for diffusive shock acceleration (DSA) of cosmic rays (CRs) can be found with a maximum proton energy as high as a few EeV. Therefore, UFOs are promising candidate sources of ultra-high-energy CRs \cite{Ehlert2024}. In addition, copious inelastic collisions of CRs with gas (pp) and radiation (p$\gamma$) are expected to take place in the UFO due to the extreme radiation field and the high gas density. Therefore, UFOs are candidate sources for both high-energy (HE) gamma-ray and neutrino.

The gamma-ray flux can escape freely for energies below $\sim100\,{\rm GeV}$, while at higher energies the UFO environment can become optically thick due to the strong radiation field associated to the AGN itself. 
A first tailored study to investigate the gamma-ray emission from UFOs was endeavored by the Fermi-LAT Collaboration \citep[][]{UFO-Fermi-LAT+Caprioli}, who carried out a stacking analysis on 11 UFOs. Their result confirmed that UFOs are gamma-ray emitters but the employed stacking technique prevents to have clear constraints on the spectrum of these sources.
Therefore, UFOs add to the list of potential gamma-ray production sites in Seyfert galaxies, where typically star forming regions, weak jets, and galaxy-scale outflows were considered.

In this work, we target NGC~4151, a Seyfert 1.5 galaxy~\citep{Osterbrock-1.5} located at $D_{\rm L}\simeq 15.8$~Mpc~\citep[][]{Yuan_distance} and known for hosting an UFO through X-ray observations. 
{The sky region of NGC~4151 is complex and potentially contaminated  by two blazars: FIRST J121134.2+390053 and 1E~1217.9+3945, located respectively at about 30 and $5$ arc-minutes from our source of interest.
Two gamma-ray sources with moderate significance (about $5 \sigma$) have been unveiled in the third and fourth data release (DR3 and DR4) of the Fermi-LAT catalog \cite{4fgl-dr3,fermilatdr4}: 4FGL~J1211.6+3901 and 4FGL~J1210.3+3928.}
{The former was associated to FIRST~J121134.2+ 390053, and the latter to the blazar 1E~1207.9+3945} {based on their spatial proximity}.
However, the {identification of 4FGL~J1210.3+3928 is challenged by the vicinity of NGC~4151, thereby making a clear identification difficult.}

{We unveil and analyze the gamma-ray emission  of 4FGL~J1210.3+3928 from 100 MeV to $\sim$100 GeV.
We perform several positional and association tests and conclude that it is not possible to clearly identify NGC~4151 or 1E~1217.9+3945 as observational counterpart.
We note an energy-dependent migration of the centroid that could be a hint of both sources emitting gamma rays. Being the source near detection threshold, only with more exposure we will be able to confirm if this tendency is real or it is a fluctuation due to the limited angular resolution of LAT combined with the small number of events. 
With the current measurements, it seems clear that both sources should be considered as possible counterparts.
In light of that, we modeled the emission of such a source considering both a blazar and UFO model (as in P23). 
Our results show that to account for the emission with a blazar model, non-physical conditions must be assumed {in the context of a one-zone leptonic model, which typically well describe BL~Lac objects such as 1E~1217.9+3945}.}
{Therefore, we interpret the detected gamma-ray flux and spectral shape in the context of the theoretical model developed in P23 and we show that the scenario of DSA at the wind termination shock of the UFO is able to explain the observations. However, we cannot exclude the possibility of blazar contamination, especially above 10 GeV.}

{The manuscript is organized as follows. In \S\ref{Section: observation} we describe the observation and the gamma-ray analysis
together with the positional tests 
and a detailed assessment on its possible blazar contamination.
Section \S\ref{Sec: Model} is devoted to the UFO model presentation and interpretation 
applied to NGC~4151.
We discuss some multi-messenger implications in \S\ref{Sec: Multim-Perspectives} and we present our conclusions in \S\ref{Sec: Conclusions}.}


\section{Fermi-LAT observation and critical study of the source counterpart}
\label{Section: observation}

\begin{figure}
    \centering
    \includegraphics[width=0.7\linewidth]{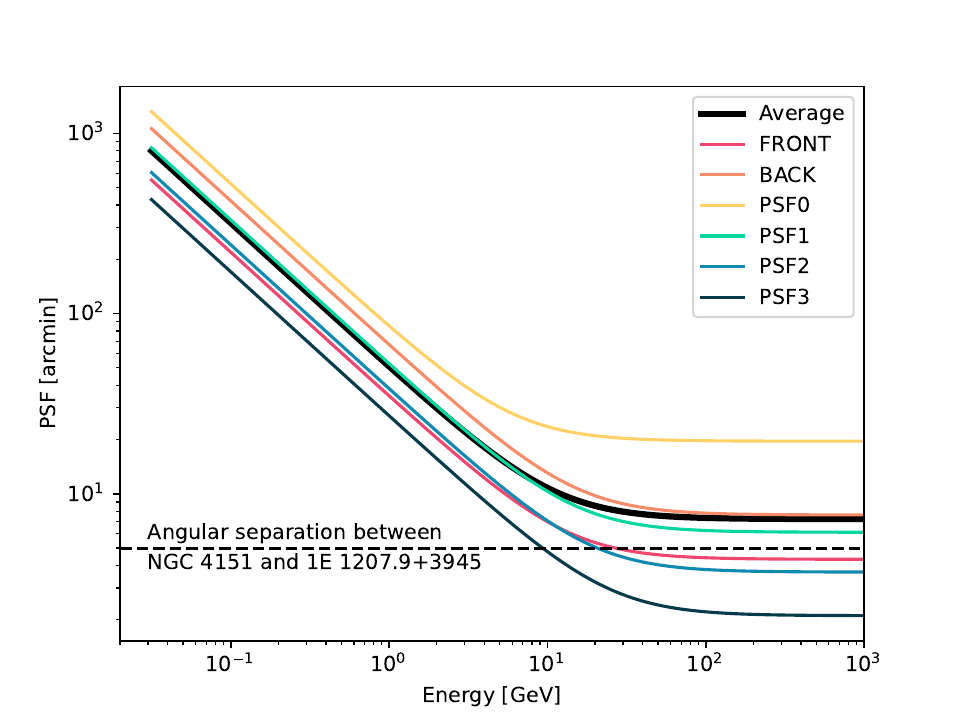}
    \caption{Point spread function of the different Fermi-LAT event selections {as indicated in the legend. They corresponds to different quartiles in the quality of the reconstruction of the event direction from the worst (PSF0) to the best (PSF3). FRONT and BACK instead corresponds to events reconstructed  at the front or at the back the tracker respectively, as this also influences the quality of the event direction reconstruction.} The black dashed line {at 5 arcmin} indicates the extent of the angular separation between NGC~4151 and the blazar 1E 1217.9+3945.} 
    \label{fig:psf}
\end{figure}

We analyze the data accumulated for more than 14 years by the Fermi-LAT in the energy range 100\,MeV\,--\,1\,TeV. {We apply the recommended data quality cuts for off-plane point-like sources, 
{\tt (DATA\_QUAL$>$0)\&\&(LAT\_CONFIG==1)},}
{and maximum zenith angle 90$^\circ$.
As the performance of the Fermi-LAT, in particular the point spread function (PSF), varies strongly with energy (see Fig. \ref{fig:psf}), we adopt different event selection criteria for the different energy ranges, following the prescription of the Fermi-LAT Collaboration described 
in \cite{4fgl-dr3}. 
This procedure allows us to select events with different PSFs \footnote{A full description of the events can be found in \href{https://fermi.gsfc.nasa.gov/ssc/data/analysis/documentation/Cicerone/Cicerone_Data/LAT_DP.html}{Fermi-LAT Cicerone}.}. 
We consider only those low-energy events (<300 MeV) characterized by the best angular reconstruction ({\tt evtype=32}, namely those of the fourth and best PSF quartile, PSF3), while at higher energies, where the PSF is more contained, all type of events are considered. 
{In Fig. \ref{fig:psf} we show the energy dependence of the Fermi-LAT point spread function for different selections as provided by the Fermi-LAT collaboration\footnote{\href{https://fermi.gsfc.nasa.gov/ssc/data/analysis/documentation/Cicerone/Cicerone\_LAT_IRFs/IRF_PSF.html}{Fermi-LAT Cicerone IRF.} }. } 
Notice that the PSF of Fermi-LAT is on average 60 arcmin between 100 MeV and 3 GeV and it is at best 10 arcmins below a few GeV.}

We perform a binned analysis on a 8$^\circ$-wide region of interest (ROI) using a nominal bin-size of 0.04$^\circ$. 
The null hypothesis of our statistical test is constructed as follows. We start from a source model (SM) comprised of all sources from the 4FGL-DR3 source catalog~\citep{4fgl-dr3}\footnote{While writing this manuscript, a new version of the Fermi-LAT catalog \citep[DR4,][based on 14 years of observations]{fermilatdr4} was announced. 
We verified that none of the newly identified sources are localized within 5 degrees from NGC~4151, thus they do not influence our analysis.} 
within 20$^{\circ}$ from the center $(l_0,b_0) = (155^\circ, 75^\circ)$ of our ROI.
\begin{figure}[h!]
\centering
\includegraphics[width=0.5\linewidth]{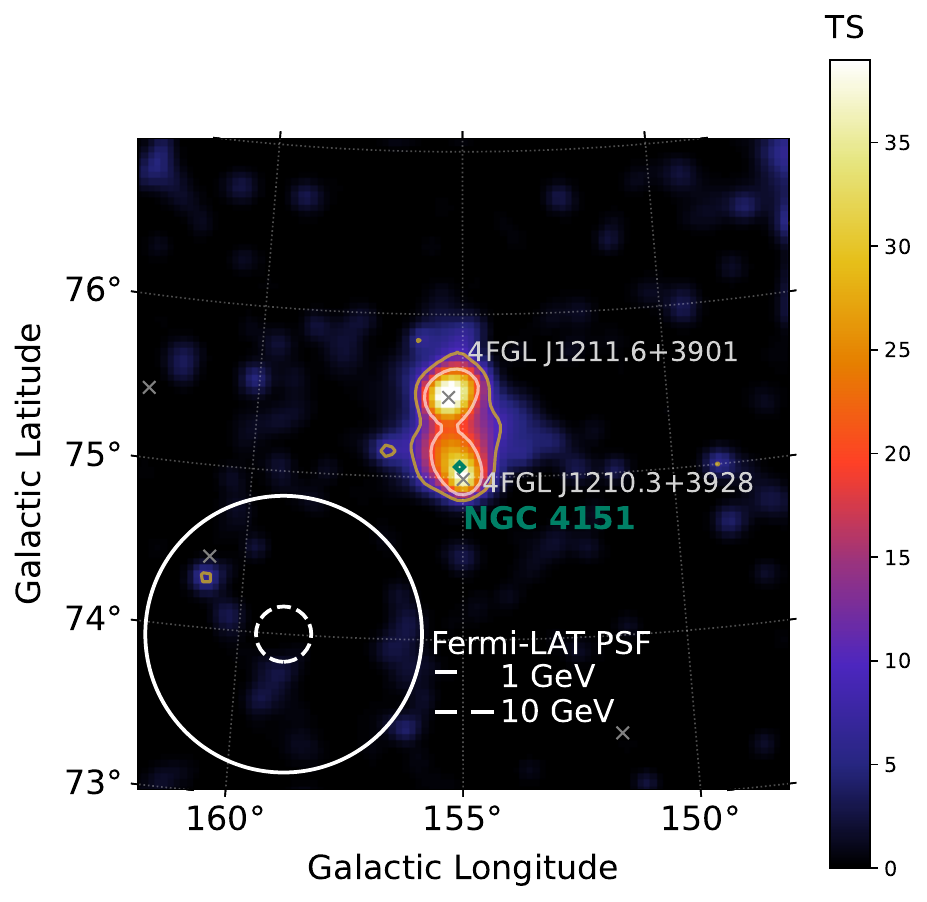}\includegraphics[width=0.5\linewidth]{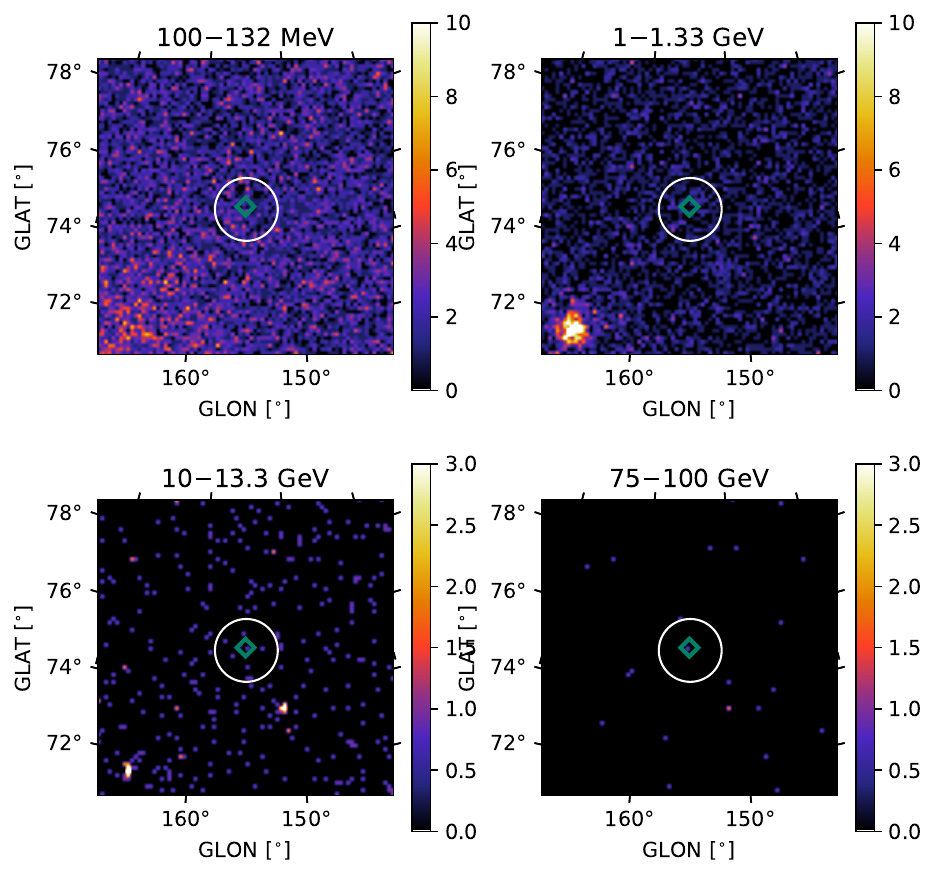}
\caption[]{{Left panel:} Test statistics (TS) map in the region of NGC\,4151 after the subtraction of all background sources except for 4FGL\,J1211.6+3901 and 4FGL\,J1210.3+3928. The 4FGL point sources are indicated as grey crosses while the position of NGC\,4151 as given in SIMBAD is indicated as a green diamond. The yellow and white contours refer to 9 and 20 TS, respectively. {Right panel: Counts map at different energy ranges over the considered ROI. The circle and the diamond are centered on NGC~4151. The size of the former indicates the extent of the PSF at 1 GeV. }
}
\label{fig:countmap}
\end{figure}
We {include in} {the SM} the Galactic ({\tt gll\_iem\_v07}) and extra-galactic ({\tt iso\_P8R3\_ SOURCE\_V3\_v1}) diffuse gamma-ray backgrounds provided by the Fermi-LAT Collaboration\footnote{\href{https://fermi.gsfc.nasa.gov/ssc/data/access/lat/BackgroundModels.html}{Fermi-LAT Background Models}}.  
We optimize the 4FGL sources in the SM on the 14-year data, using catalog parameters as initial seed and we fit the parameters of the diffuse components. 

We find two 4FGL sources near the location of NGC~4151\footnote{\href{http://simbad.u-strasbg.fr/simbad/sim-id?Ident=NGC++4151}{Simbad catalog} coordinates of NGC~4151: $(l,b) = (155.08,75.06)^\circ$} and within the PSF of the LAT instrument: 4FGL\,J1210.3+3928 (hereafter referred to as SRC-1) and 4FGL\,J1211.6+3901 (hereafter referred to as SRC-2).   
Within the Fermi-LAT catalog, both sources are associated to blazars, due to their spatial coincidence with 1E\,1207.9+3945 and FIRST\,J121134.2+390053, respectively. However, the former is located within 5 arc-minutes from NGC~4151 making a clear separation of the two sources extremely challenging given the extension of the Fermi-LAT PSF.
After the optimization of the background sources, we search for emission in the direction of NGC~4151\footnote{An indication for gamma-ray emission (4.2\,$\sigma$) coincident with the position of NGC\,4151 was already reported in the previous analysis of \cite{UFO-Fermi-LAT+Caprioli}, based on data with a 11-year-long exposure.} {, by removing and remodeling SRC-1 and SRC-2}. 
{Fig.~\ref{fig:countmap} illustrates the selected ROI in terms of test statistics (TS) maps after the subtraction of all background sources except for the two central ones. Here, the gray crosses represent the Fermi-LAT sources and the green diamond indicates the location of NGC~4151.}
{Two hotspots emerge, emphasized in the TS map by yellow contours. The northern one well matches the position of 4FGL~J1211.6+3901, while the southern peak coincides with the position of NGC~4151.}

We re-model the two sources by fitting their position and extensions.
We find the following best-fit positions: SRC-1 $(l_*,b_*)=(154.91 \pm 0.04, 75.02 \pm 0.03)^\circ$ and SRC-2 $(l_{\smallstar},b_{\smallstar})=(155.28 \pm 0.03, 75.49 \pm 0.02)^\circ$.  
{Therefore, the two sources are well separated, and SRC-2 can be considered as a background source for SRC-1. However,}
we discuss in the following subsections the properties and the potential impact of SRC-2 on the spectrum of SRC-1. 

\begin{figure}
\centering
\includegraphics[width=1.0\columnwidth]{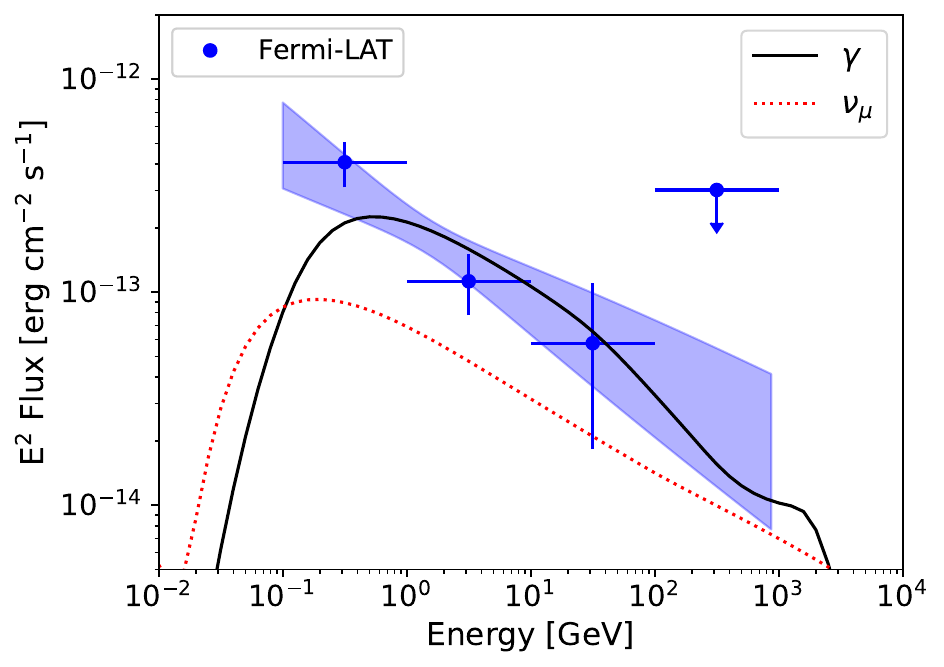}
\caption{Observed gamma-ray flux compared with the multimessenger model prediction. Data points (cyan) are shown along with the 1\,$\sigma$ uncertainty band. The thick black line represents the predicted gamma-ray flux while the dotted red curve is the associated per-flavor neutrino flux.}\label{fig:Image_1}
\end{figure}

We model the emission of SRC-1 as a point-like source, centered on $l_*,b_*$, although the exact position of the center, within limits, does not affect the SED, as shown later.  We assume a power-law spectrum $N_0 (E/E_0)^{-\alpha}$ normalized at the pivot energy $E_0=1$\,GeV. 
The obtained spectral energy distribution (SED) is displayed in Fig. \ref{fig:Image_1}. 
The resulting significance is 5.52\,$\sigma$ with best-fit normalization $N_0=(1.3\pm 0.2)\cdot 10^{-10}\,{\rm GeV}^{-1}\,{\rm cm}^{-2}\,{\rm s}^{-1}$ and spectral index $\alpha= 2.39 \pm 0.18$. 
The background source SRC-2 is found with a power-law index $\alpha_{\rm SRC-2}=1.75 \pm 0.179$ and normalization of $N_{0, \rm SRC-2}=4.91 \pm 2.29 \times 10^{-11}$ GeV$^{-1}$ cm$^{-2}$ s$^{-1}$.

\begin{figure}[h]
    \centering
    \includegraphics[width=0.8\linewidth]{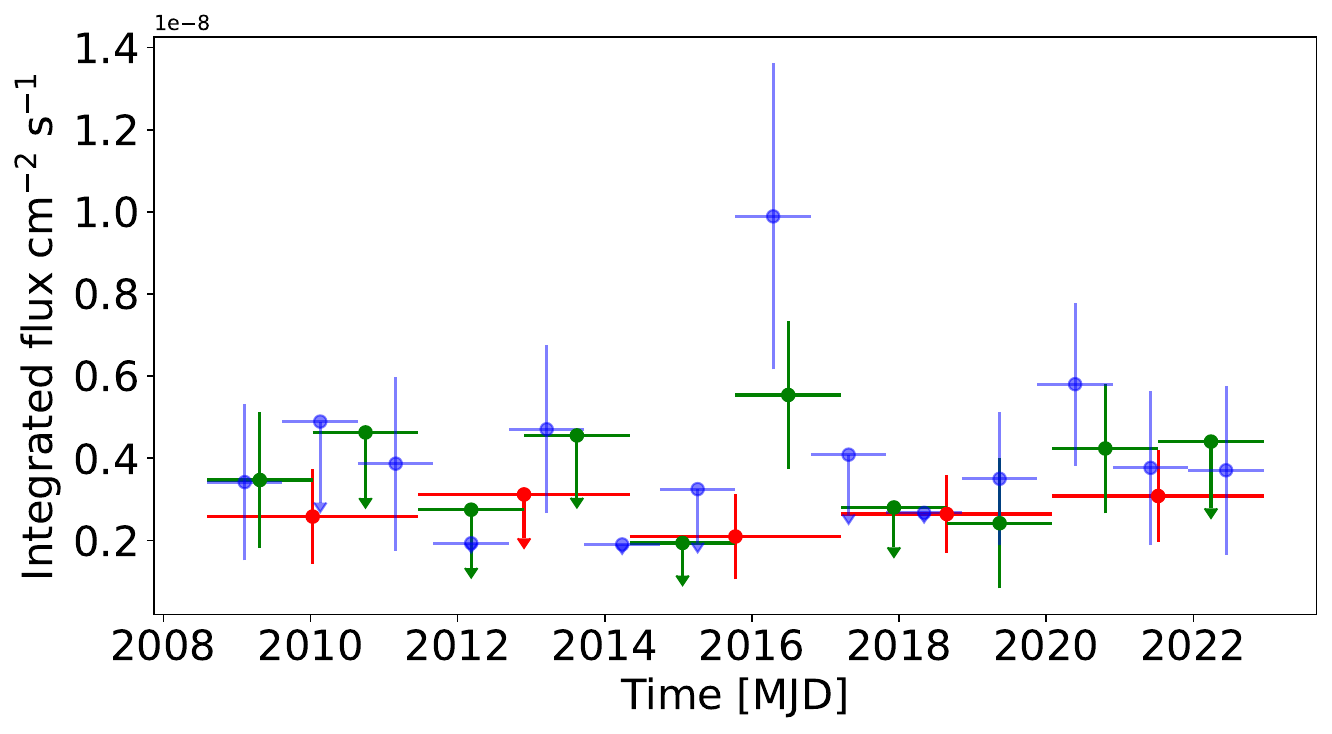}
    \caption{Gamma-ray lightcurve derived by Fermi-LAT for the source SRC-1 in different temporal bins of 1 year (blue), 1.4 years (green) and 2.9 years (red). }
    \label{fig:Fermi-LAT lightcurve}
\end{figure}

Finally, we explore the temporal variation of the gamma-ray flux to check if there is any hint of time-dependence of SRC-1. Given the low significance of the source we conclude that there are no indications suggesting a flaring activity typical of blazars for SRC-1 as displayed in Fig. \ref{fig:Fermi-LAT lightcurve}.

The gamma-ray luminosity of SRC-1 in the energy band $0.1-100$\,GeV is $L_{\gamma} \simeq3.7 \cdot 10^{40} \, \rm erg \, s^{-1}$, which is only a fraction $\sim 0.04 \%$ of the bolometric luminosity of NGC~4151, $L_{\rm bol}\simeq10^{44} \, \rm erg \, s^{-1}$~\citep{Bolometric}.
In what follows, we critically assess the emission of SRC-1 by testing the spatial and spectral correspondence of its emission with the two possible physical counterparts: NGC~4151 and the blazar 1E~1207.9+3495.

\subsection{4FGL J1210.3+3928 (SRC-1): identification and {localization} }
\label{app:identification}

Two possible counterparts are valuable candidates for SRC-1: NGC~4151, and the blazar 1E\,1207.9+3945, a BL Lac at redshift z=0.62~\cite{Morris91}. 
Given the low significance of the emission, a separation into two sources is not possible. When trying to include two sources in the model, the difference in log-likelihood is $\sim$ 2, indicating that the two sources are indistinguishable and it is not possible to obtain two separate SEDs. 
{In fact, if both NGC 4151 and 1E 1207.9+3945 are assumed to be present simultaneously, the one postulated as a background source will inevitably overwhelm the other.
This would result in the spectrum of the second source to be characterized by only upper limits as discussed in \cite{Murase_Seyfert}.}
Nevertheless, we performed different tests to evaluate whether one counterpart dominates the emission, both considering the localization and the spectral distribution.

\begin{figure}
    \centering
   
    \includegraphics[width=0.3\linewidth]{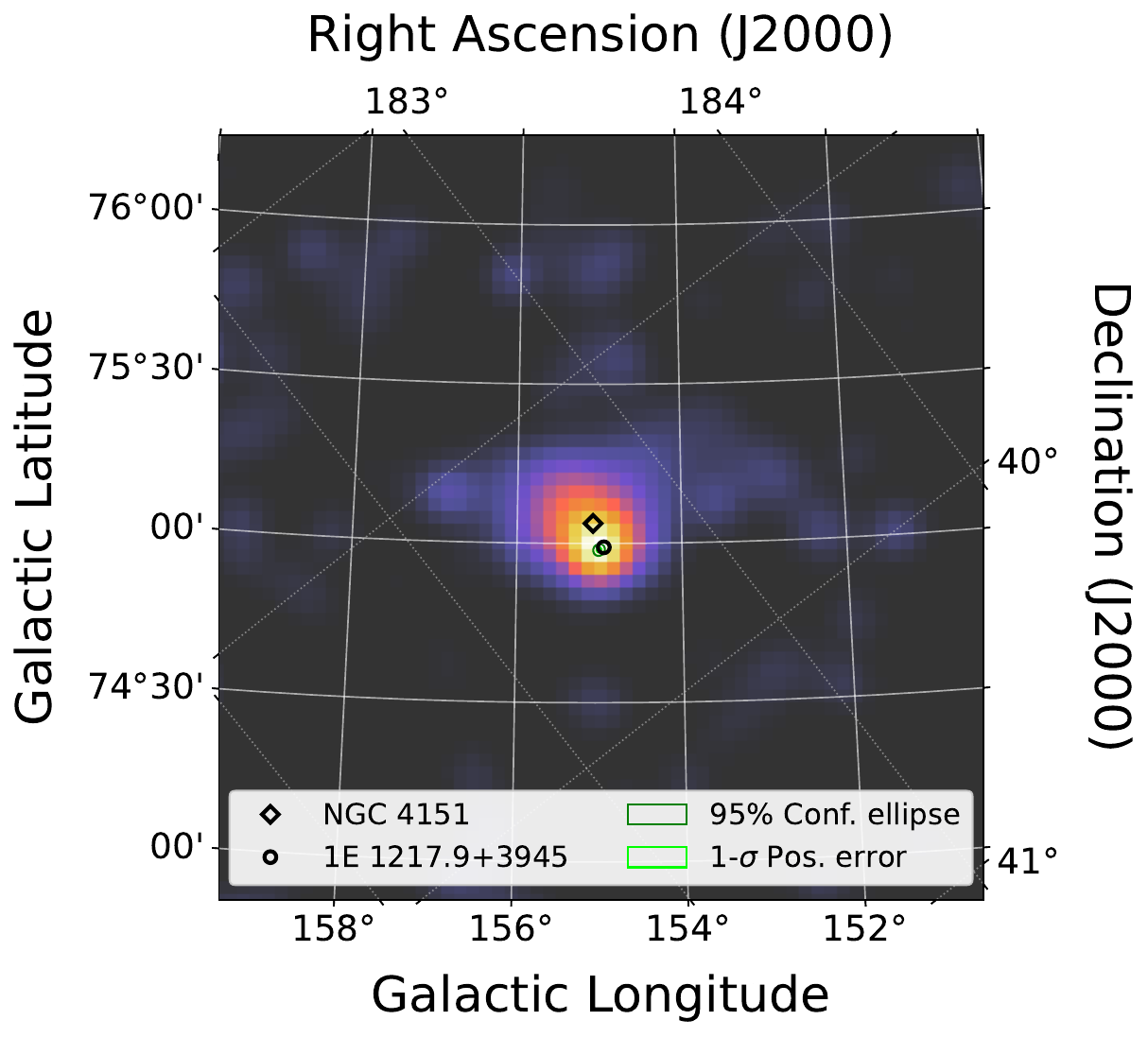}\includegraphics[width=0.3\linewidth]{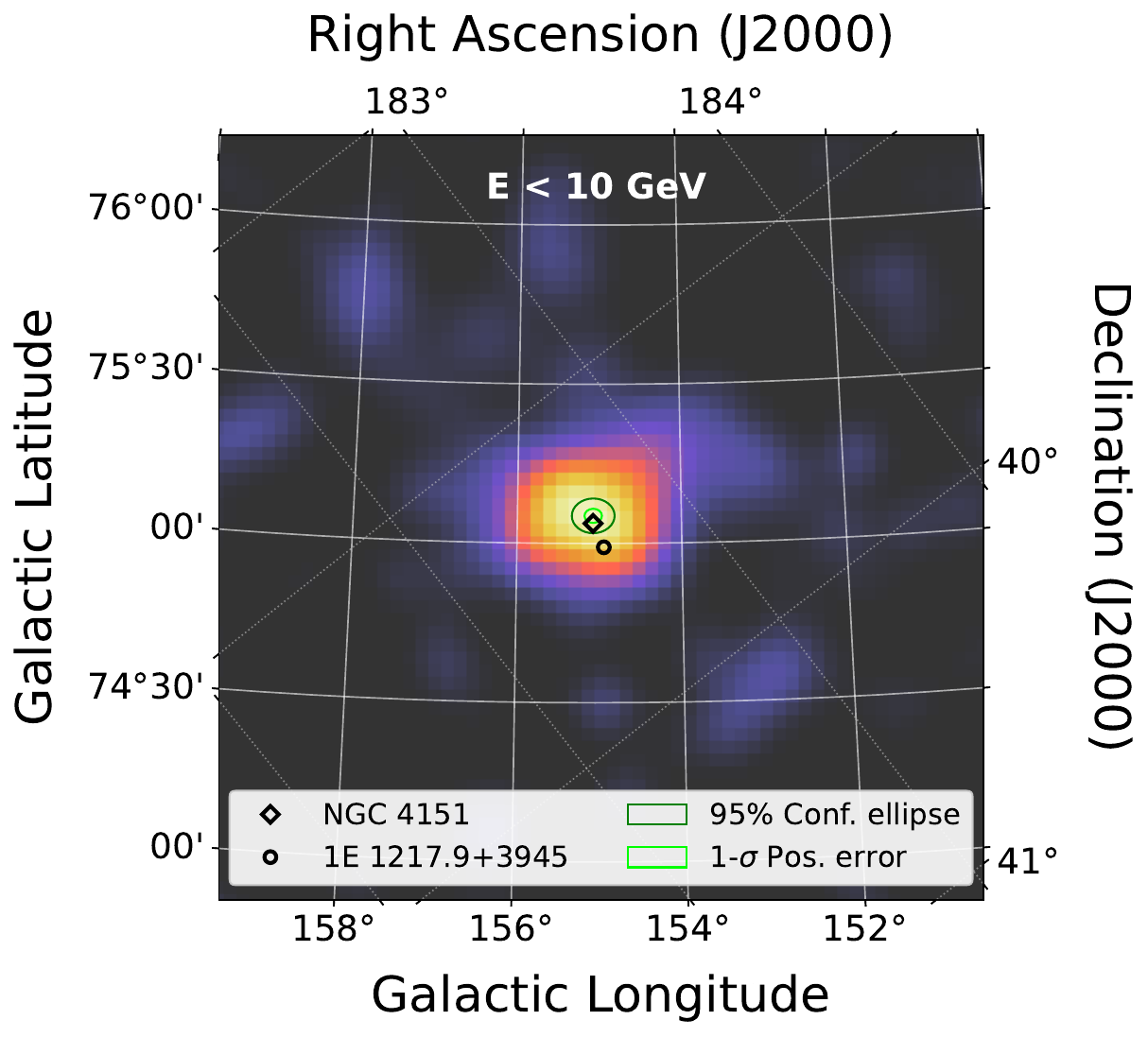}
    \includegraphics[width=0.3\linewidth]{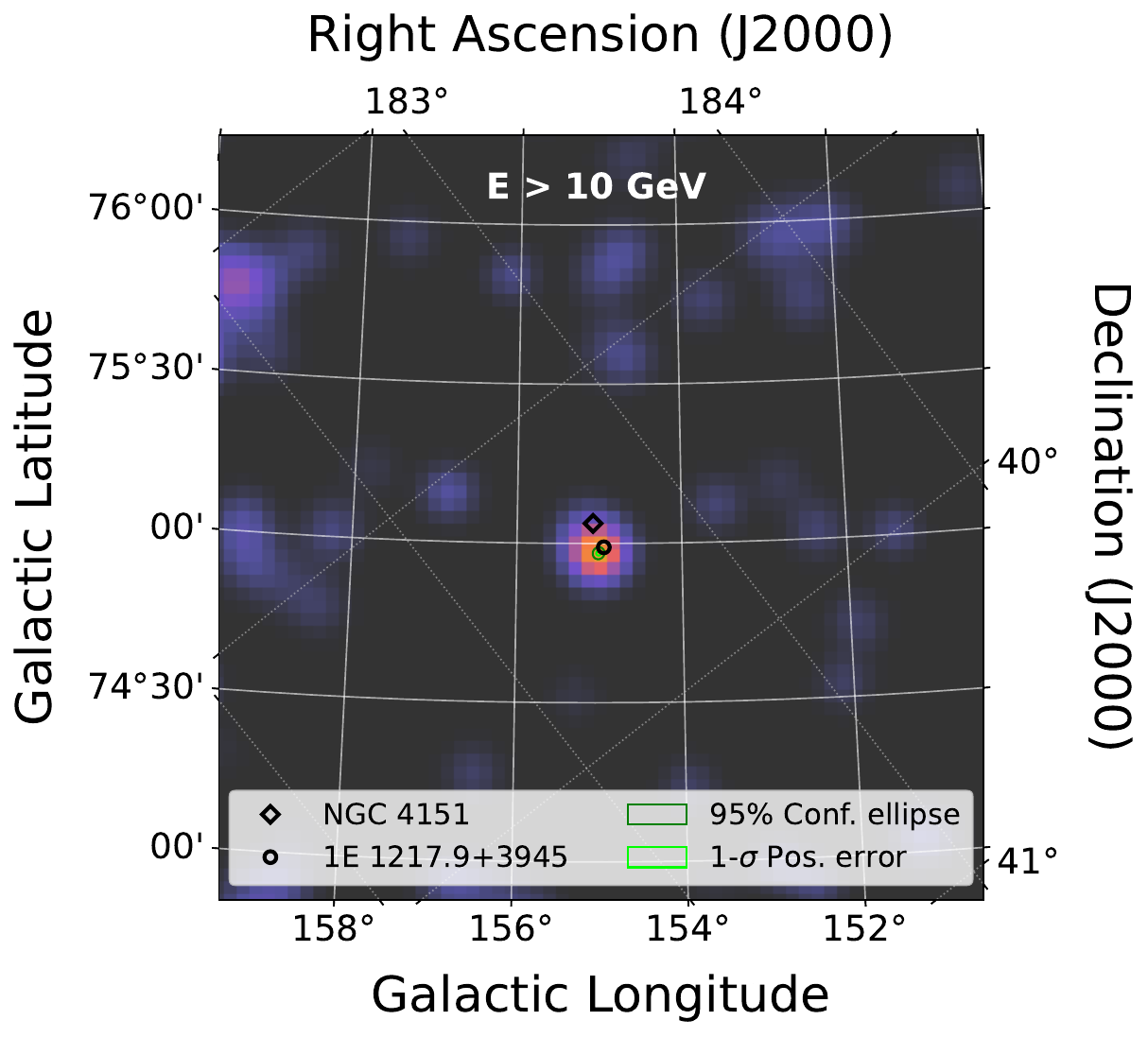}
    \caption{Test statistics maps obtained for different energy ranges: the full energy range (100 MeV-1 TeV) is shown in the leftmost panel, while the center and in the rightmost panels illustrate the results obtained below and above 10 GeV respectively. SRC-1 is always excluded from the model. In all maps the black diamond and the black circle indicate the position of NGC~4151 and 1E~1217.9+3945. The green ellipses indicate the result of the localization tests for SRC-1: the 1 sigma uncertainty is indicated in light green, while the 95\% confidence ellipse is drawn in dark green.}
    \label{fig:TS-E-dependent}
\end{figure}

\begin{figure}[h!]
    \centering
    \includegraphics[width=\textwidth]{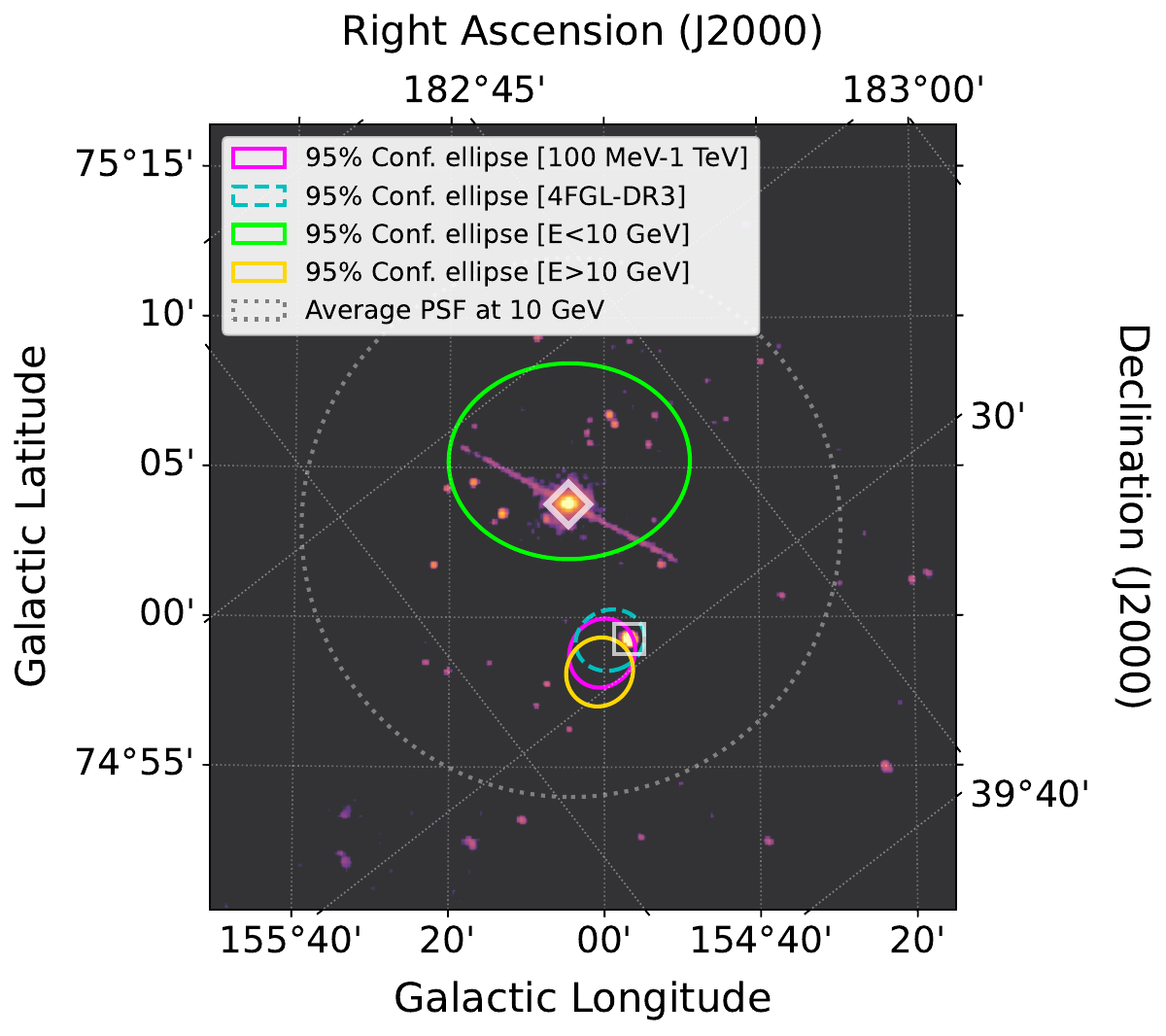}
    \caption{95\% confidence ellipses obtained in the different configuration as explained in the legend and text. Different energy ranges are compared. The confidence ellipses are over-plotted to the Chandra/ACIS image of NGC~4151, obtained in the keV energy range. The position of the latter is indicated as a white diamond, while the position of the nearby blazar is indicated as a white square. The grey dashed circle indicate the Fermi average PSF at 10 GeV.}
    \label{fig:pos_newtests1}
\end{figure}

We first run a localization test, using the Fermi-tools.  
We repeat this test for different energy ranges in order to prove if the source location is stable in energy.  
The results are shown in Table \ref{tab:locs} where we report the best-fit positions of the centroid, together with their relative 68\% ($r_{68}$) and 95\%  ($r_{95}$) uncertainties. 
These correspond to the geometric mean values of the major and minor axes of the 68\% and 95\% confidence error ellipses \citep{fermi1FGL}. 
Figure~\ref{fig:TS-E-dependent} illustrates the dependence on the energy range of the TS-map. We consider the full range (left panel), the low-energy range (E<10 GeV, central panel) and the high-energy range (E>10 GeV). 
In Figure~\ref{fig:pos_newtests1} we show the fitted position and their relative confidence ellipses obtained in the different energy bands overlaid, respectively, with the Fermi-LAT TS map and with a map of the region obtained by Chandra/ACIS in the keV range\footnote{{Data obtained from the Chandra Data Archive (obsID: 348 (PI: Wilson, exposure 32.4 ksec)}}, where the two candidate counterparts are easily distinguishable. 
As one can see from the table and the image, these localization tests attempt to speculate on the presence of a low-energy counterpart towards NGC~4151, while, at higher energies, a contamination from the nearby blazar could be present.
This hypothesis is further supported by a spatial shift of the centroid towards the position of 1E~1217.9+3945. 
However, since both sources fall within the angular resolution of Fermi-LAT, a definitive distinction cannot be established based solely on positional argument. 
Therefore, in section \ref{Sec: Model} we complement the present discussion with a physical modeling of both possible counterparts.

\begin{table}
    \centering
    \begin{tabular}{lccc}
    \hline
         &   Best-fit (l,b)$^\circ$& $r_{68}$ ($^\circ$)&$r_{95}$ ($^\circ$)\\
         \hline
 \multicolumn{4}{c}{Bin=0.04$^\circ$}\\
\hline 
Full range & (155.0046 ,74.9802) &0.022 & 0.037\\
< 10 GeV &(155.0753,75.0870) & 0.076& 0.1209 \\
>10 GeV & (155.010,74.979) & 0.023& 0.038 \\

\hline
\end{tabular}

    \caption{Result of localization fitting in the different energy ranges.}
    \label{tab:locs}
\end{table}

\begin{figure}
     \includegraphics[width=7.5 cm,height=5.5 cm]{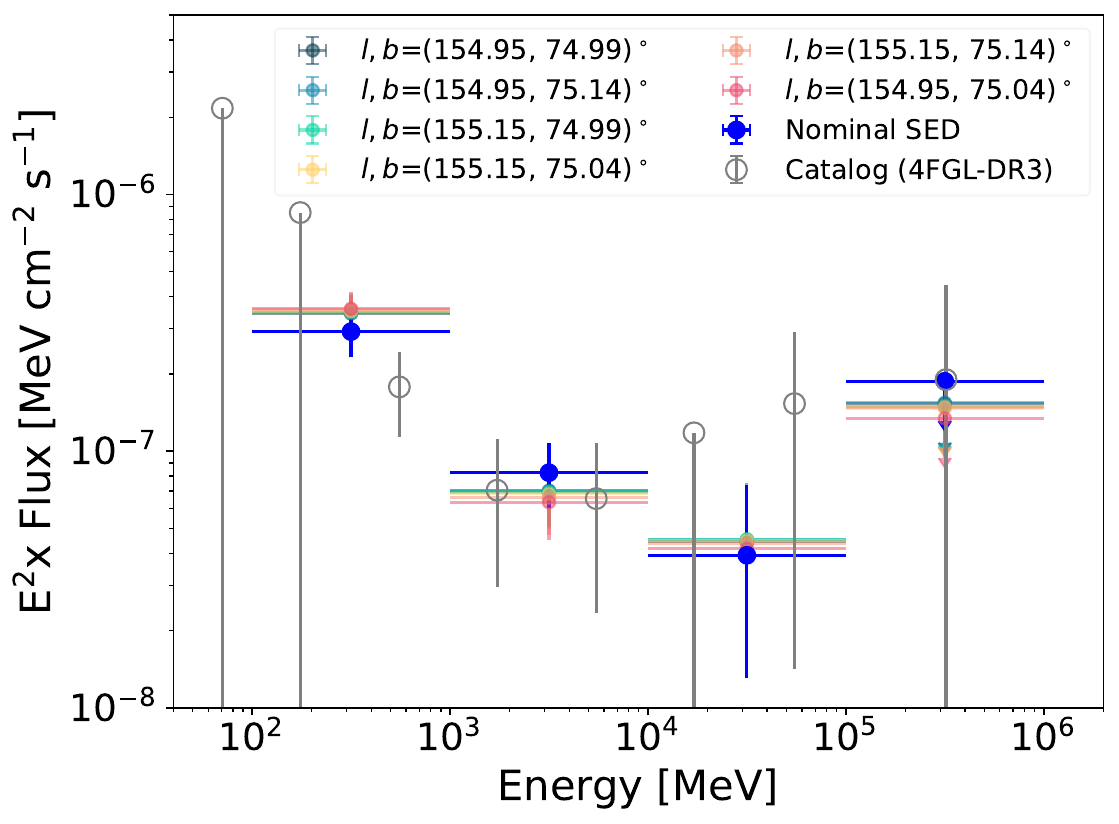}\includegraphics[width=7.5 cm,height=5.5 cm]{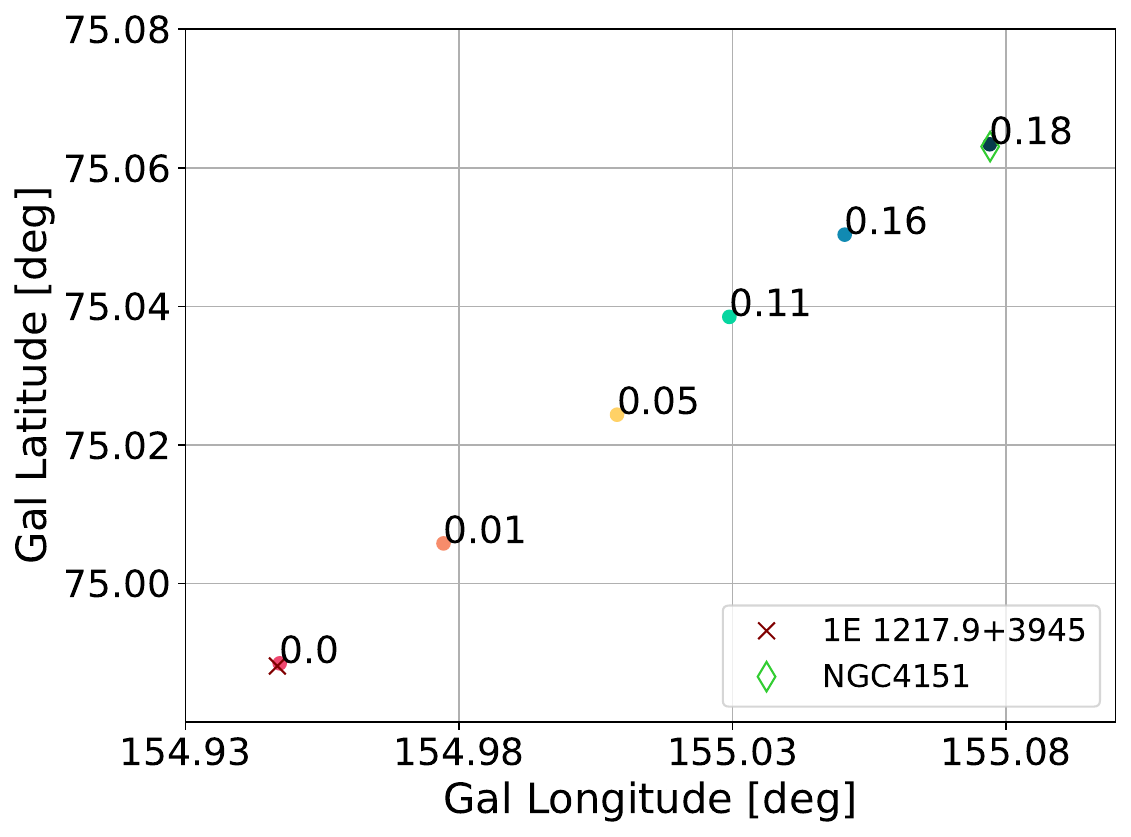}
    \caption{{Left panel: SED of SRC-1 obtained for different center positions as indicated in the legend. The blue circles and the gray empty circles indicate the nominal SED, presented in Fig. \ref{fig:Image_1}, and the SED reported in the Fermi 4FGL catalog \citep{4fgl-dr3}, respectively. Right panel: difference in logarithmic likelihood obtained in the global fit when displacing SRC-1 from the blazar coordinates (red cross) to the coordinates of NGC~4151 (green diamond). }}
    \label{fig:localization_test}
\end{figure}

It should be noted that the SED extracted from SRC-1 does not vary within the PSF. 
We test this by varying the position of the centroid from the center of NGC~4151 to the center of 1E~1217.9+3945. 
The SED of the source does not vary with the choice of the centroid, as shown in the left panel of Figure \ref{fig:localization_test}. 
This is also reflected in the likelihood of the different positions as displayed in the right panel. Here, we report the resulting likelihood values for the different positions moving from NGC~4151 to the blazar. We see that the likelihood is slightly larger towards NGC~4151, with respect to the blazar, however, the difference is too small to allow a robust identification of a spatial counterpart.

\subsubsection{Influence of 4FGL\,J1211.6+3901 (SRC-2)}
\label{app: lat supplements}

\begin{table}[t]
\centering             
\caption{Significance and spectral parameters of SRC-1 from different source models (SMs) as explained in the main text. The flux normalization $N_0$ at the pivot energy $E_0 =  1$~GeV is in unit of $10^{-10}$ GeV$^{-1}$ cm$^{-2}$  s$^{-1}$.}
\vspace{0.5cm}
\label{tab:tests}
\renewcommand{\arraystretch}{1}
\linespread{1}\selectfont
\begin{tabular}{c|ccccc}
\hline 
\makebox[1.cm][c]{energy range}& \makebox[0.8 cm][c]{SM} & \makebox[0.8 cm][c]{AIC} & \makebox[0.8 cm][c]{$\sigma_{\rm SRC-1}$}& \makebox[1.cm][c]{$N_0$}  & \makebox[1. cm][c]{$\alpha$}\\
\hline
\multirow{4}{*}{> 100 MeV} & {SM$_{11}$}  & {47120.57}  & {5.46}  & {1.3\,$\pm$\,0.3}    & {2.42\,$\pm$\,0.2} \\
                           & SM$_{10}$ & 47153.50  &  6.57 & 1.6\,$\pm$\,0.3 & 2.48\,$\pm$\,0.18\\
                           &  SM$_{01}$ &47146.45   & -- & -- &-- \\
                           &  SM$_{00}$ & 47192.69 &-- &-- & -- \\
\hline
 \multirow{4}{*}{> 1 GeV}  &  SM$_{11}$  & 55732.33 & 4.53   & 0.9\,$\pm$\,0.4 & 2.18\,$\pm$\,0.23\\
                           &  SM$_{10}$  & 55760.94  & 5.20 & 1.1\,$\pm$\,0.4  & 2.20 \,$\pm$\,0.22 \\
                           &  SM$_{01}$  & 55744.32  & -- & -- &-- \\
                           &  SM$_{00}$ & 55783.43  &-- & -- & -- \\
\hline
\end{tabular}
\end{table}

\begin{figure}[h!]
\centering
\includegraphics[width=0.7\linewidth]{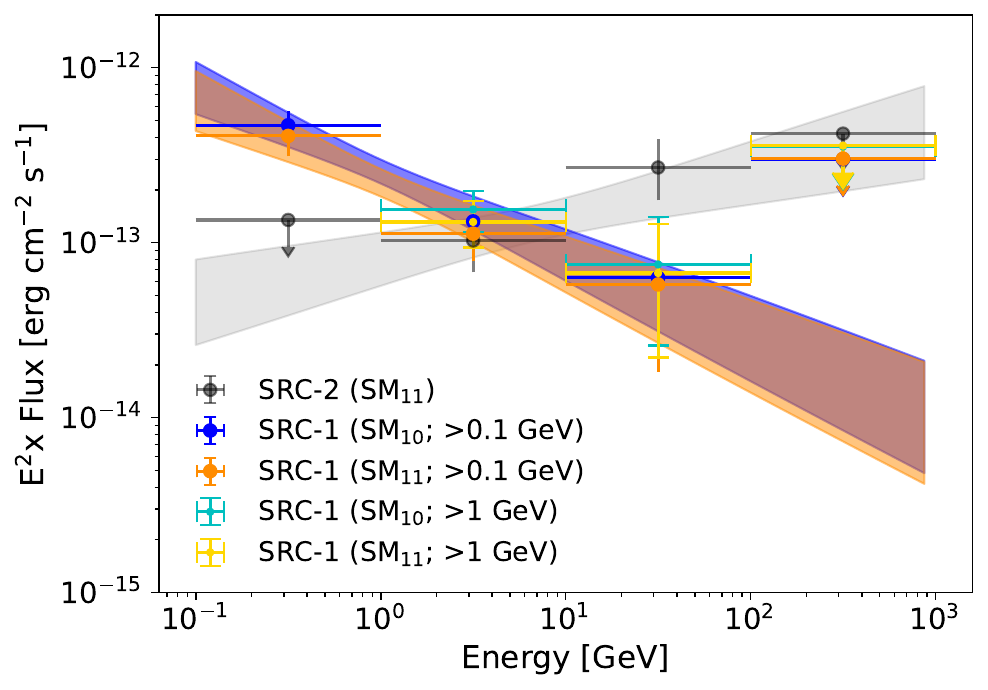}
\caption{The spectral energy distribution of the blazar 4FGL\,J1211.6+3901 (SRC-2) and of NGC\,4151 (SRC-1) in the different configurations.}\label{fig:sed_tests}
\end{figure}

We run additional tests to assess the quality of our detection of SRC-1 due to the possible influence of the nearby SRC-2. 
We proceed with a detailed likelihood analysis aimed at investigating whether the SED could be influenced. 
To do that, we construct four SMs, all including the optimized background sources but with different assumptions for the configuration of the central excess. 
They are defined as follows: SM$_{00}$ excludes both SRC-1 and SRC-2, SM$_{10}$ includes only SRC-1, SM$_{01}$ includes only SRC-2 and SM$_{11}$ includes both sources. 
We compute the likelihood of each SM, and evaluate the significance and spectral shape of SRC-1 in the different configurations. 
To evaluate which SM is preferred, we use the Akaike information criterion and computed ${\rm AIC} = 2k - 2\mathcal{L}$, where $k$ is the number of parameters of the model and $\mathcal{L}$ is the log-likelihood that results from the fit. 
The model favored by the lower AIC is SM$_{11}$ (highlighted in Table \ref{tab:tests}), namely the SM including both sources. 
We compute the significance of SRC-1 as $\sigma_{\rm SRC-1} = \sqrt{2(\mathcal{L}_{1j}-\mathcal{L}_{0j})}$, where $\mathcal{L}_{ij}$ are the log-likelihood of the respective SM$_{ij}$. 
In the latter we account for the situation where SRC-2 is included or not in the SM, and for both cases we extract the spectral energy distribution and compute the spectral parameters.

We perform this test both with gamma-rays of energies $>100$\,MeV and with only those of energies $>1$\,GeV. 
At higher energy, the Fermi-LAT point spread function significantly improves, maximizing the chance for clear separation of the two sources. 
The resulting spectral parameters and significance are reported in Table~\ref{tab:tests} along with the AIC values. 
The significance in all cases is $\gtrsim 5\,\sigma$, confirming the detection of SRC-1 despite the vicinity of SRC-2. 
The spectral energy distributions for the different configurations are reported in Fig.~\ref{fig:sed_tests}. The spectrum is stable disregarding of SRC-2, which, as said, is modeled as a power-law of index $\alpha_{\rm SRC-2}=1.75 \pm 0.179$ and normalization of $N_{0, \rm SRC-2}=4.91 \pm 2.29 \times 10^{-11}$ GeV$^{-1}$ cm$^{-2}$ s$^{-1}$.

\subsubsection{Correlation}

\begin{figure}[h!]
    \centering
    \includegraphics[width=0.7\linewidth]{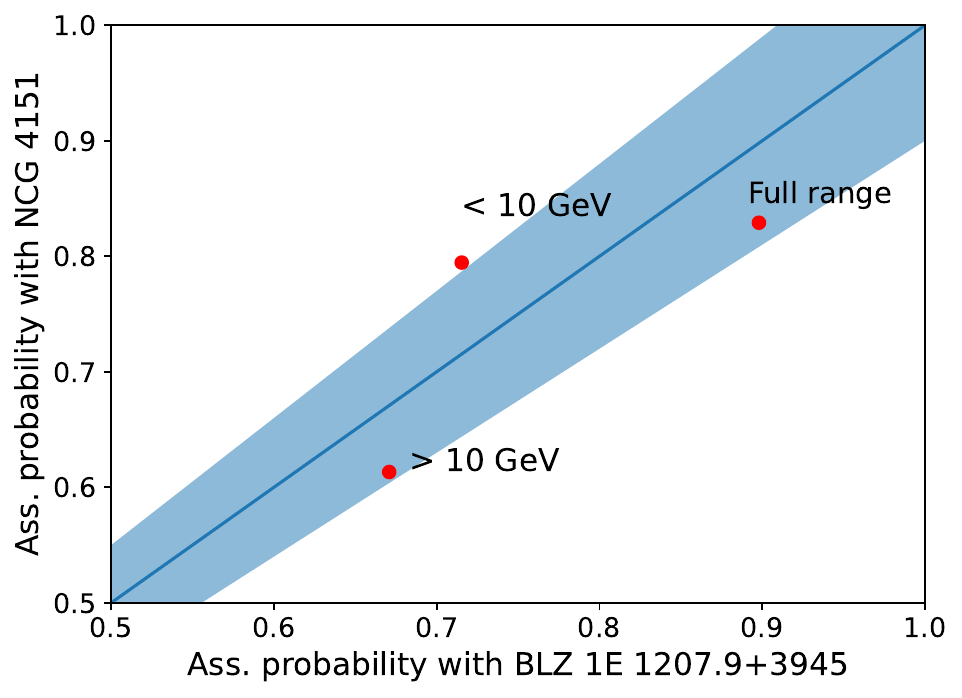}
    \caption{{Association probability resulting from identification tests in the different energy domains as explained in the text.}}
    \label{fig:prob}
\end{figure}

We further explore the correlation of the source with the proposed counterparts using the Bayesian association method described in \cite{Abdo2FGL} and implemented in \texttt{gtsrcid} of the Fermi science tools.  
This method evaluates the association by weighting the source-counterpart distance to the positional uncertainties and to the density of similar objects in the surroundings. 
We considered the BZCAT \citep{Massaro2015} for blazars and the BAT hard-source catalog for Seyfert Galaxies like NGC 4151. 
The probabilities are evaluated at different energies, as argued earlier, and the results are displayed in Figure~\ref{fig:prob}. 
{One can see that according to the energy range, one or the other counterpart is favored, but none emerges clearly from this test. 
Considering that the differences in probabilities are smaller than 10\% (blue band in the figure), we conclude that both sources are reasonable candidates. This is consistent with what we find in the localization tests.}

{Based on the present results and those on the localization, we argue that the only robust way to perform a reliable source identification is through a multi-wavelength spectral interpretation, as performed in the following.}

\subsection{Spectral evaluation: testing the blazar hypotesis}
\label{Subsec: Blazar}

\begin{figure}[h!]
    \centering
    \includegraphics[width=0.7\linewidth]{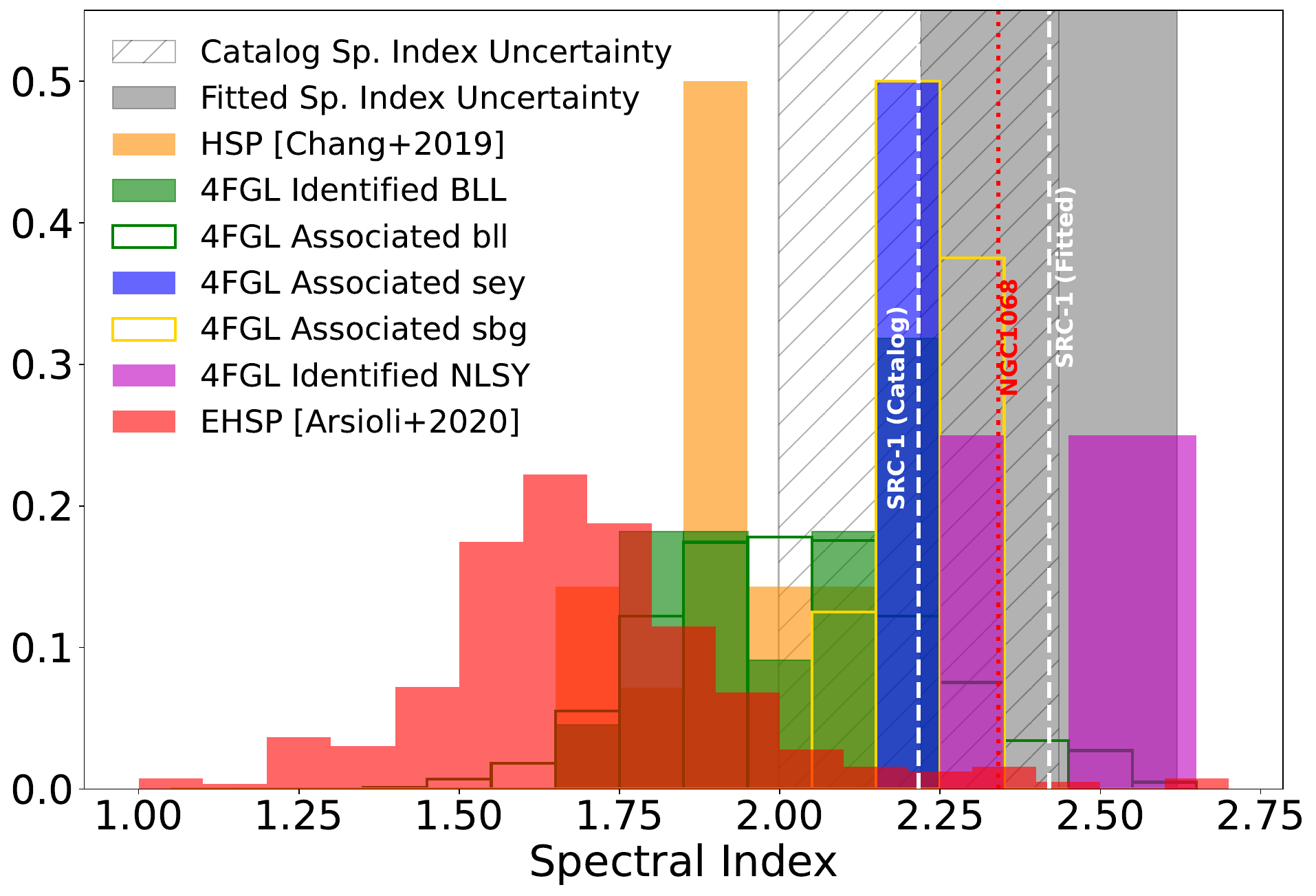}
    \caption{Distribution of spectral indexes of different source populations as reported in the 4FGL source catalog \citep{4fgl-dr3}. The fitted spectral index of SRC-1 is also plotted as reported by the catalog and by our analysis (dashed white lines) along with its associated uncertainty (grey areas). The spectral index of NGC 1068 is also indicated in red.}
    \label{fig:stats}
\end{figure}
{As NGC~4151 and 1E~1207.9+3945 are too close to allow a successful localization test 
of the true counterpart of SRC-1, we investigate here whether the derived SED can be interpreted as emission from the latter. Before doing so, we remind the reader that the source spectrum does not vary within the considered localization (Fig. \ref{fig:localization_test}). } 

\begin{figure}[h]
      \centering
      \includegraphics[width=0.8 \linewidth]{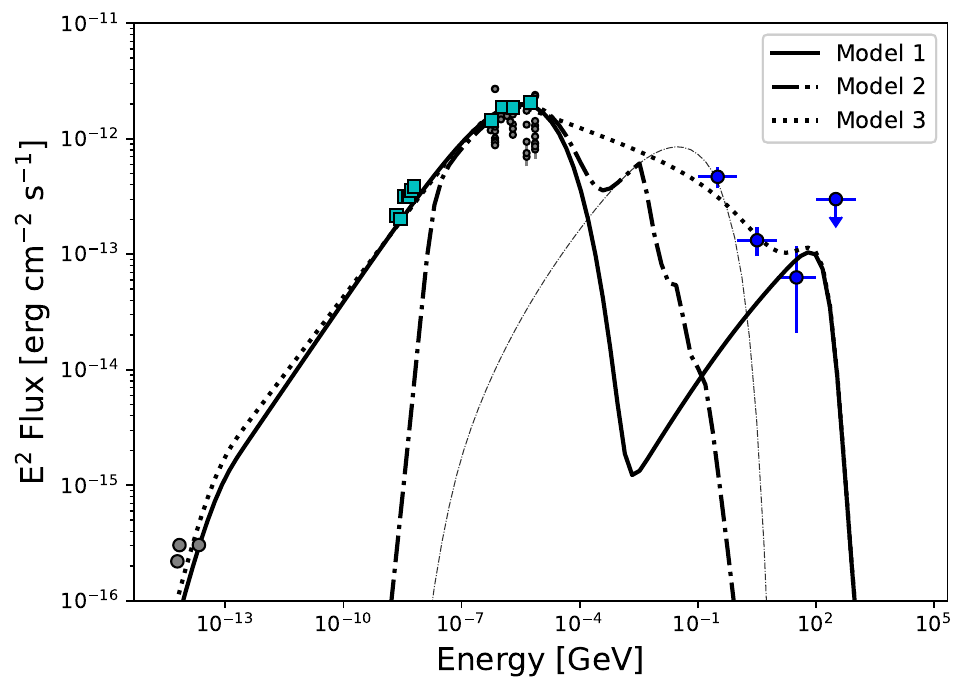}
      \caption{Multiwavelength SSC modeling of the blazar 1E~1207.9+3945. The three thick lines represent the three alternative models, while the thin dot-dashed line represent the un-absorbed SSC component of Model 2. The cyan data points are from the analysis of \cite{Maselli2008}  of Swift UVOT and XRT data. The radio data (grey circles) come from VLA observations \citep[][]{radio1985}, while grey points are other asynchronous X-ray observations from Swift and XMM-Newton. The blue points are the Fermi-LAT data points collected in this work.}
    \label{fig:sed_blazar}
\end{figure}

We start by comparing in Figure~\ref{fig:stats} the spectral index of SRC-1 with the spectral indexes of the identified BL Lacs in the 4FGL catalog. 
As one can see from the histogram, the soft LAT spectrum would make this source an outlier among the blazar population. 
The association of the LAT SRC-1 with 1E~1207.9+3945 is even more problematic if one qualitatively observes at its multi-wavelength emission (Fig.~\ref{fig:sed_blazar}). 
In fact, the SED of 1E~1207.9+3945 shows a synchrotron (SYN) peak at around $10^{18}$ Hz, classifying it as an extremely-high-frequency-peaked-BL Lac (EHBL) \citep{Foffano19}. 
An EHBL with a second SED peak below 100 MeV is extremely rare, and this kind of SED can be easily ruled out from a theoretical point of view. Assuming that the emission is primarily leptonic and due to synchrotron-self-Compton (SSC) scattering as typical in HBLs, we can use the analytical formulae derived in \cite{Tavecchio98} to put constraints on the model parameters. 
From the equation relating the peaks of the SED component, assuming $\nu_{\rm syn} = 10^{18}$ Hz, and $\nu_{\rm ssc} = 10^{22.8}$ Hz, the estimate of the magnetic field in the emitting region is $B = 9.1\times10^5 (\delta/10)^{-1}\ \textrm{G}$ (with $\delta$ the Doppler factor of the emitting region), which is already an unrealistic value, several orders of magnitude larger than typical values inferred for HBLs. 
On the other hand, the equation relating the peak luminosities results in a relatively standard amplitude of $B = 0.04\ (\delta/10)^{-3} (\tau/1 \, \rm{month})^{-1} \textrm{G}$ (with $\tau$ the observed variability time-scale). 
It is impossible to reconcile these vastly different values. 

This simple analytical estimate allows us to disfavor 1E~1207.9+3945 as the only counterpart of SRC-1. 
It is then legitimate to ask what can be the maximum contribution of the source to the gamma-ray signal. 
With this goal, we model the SED of the source in a leptonic framework (using the code described in \cite{Cerruti15}), assuming model parameters typical for extreme HBLs: $\delta=50$, $B = 1$ mG, $R=3.2\times10^{17}$ cm, where $R$ parametrizes the location of the emitting region. Electrons are parametrized as a power-law distribution with spectral index $\alpha=2$ between $\gamma_{\rm min}=300$ and $\gamma_{\rm max}=3\times10^6$.

The associated SYN+SSC spectrum is shown in Figure \ref{fig:sed_blazar} as Model 1 (solid line) and it is compared with multiwavelength data collected by several instruments.
The modeling is performed on data collected simultaneously by Swift UVOT and XRT (green square), as reported in \citep[][]{Maselli2008}. Additional flux points (grey dots) collected respectively by VLA, Swift and XMM-Newton and retrieved from the SED builder platform are included for comparison. 
In order to avoid source confusion, we exclude data from other X-ray observatories that have an angular resolution bigger than the separation ($\sim$ 5 arcmin) between the blazar and NGC 4151.
Given that there is no firm gamma-ray detection from the source, the model is degenerate, and we present here a solution that maximizes the emission in the LAT band without violating it. 
As it can be seen, 1E~1207.9+3945 could contribute and possibly dominate the emission of SRC-1 only above 10 GeV, while the bulk of the LAT signal with its soft spectral index, can be confidently attributed to NGC 4151.

In order to explore the possible contamination of 1E~1207.9+3945 to the SRC-1 spectrum below 10 GeV we develop two alternative models. 
In particular, Model 2 (dot-dashed line) and Model 3 (dotted line) are attempts to explain the low-energy LAT data respectively with their SSC and SYN components. 

In Model 2, we force the peak of the SSC component in the MeV range. 
This is only possible by assuming a very low value of the variability timescale, and thus of R. The adopted parameters are: $\delta = 10$, $B = 9.1 \times 10^5$ G, and, most importantly, $R = 1.7 \times 10^{10}$ cm, which is smaller than the solar radius, and represents the most important issue with this model. It is clear from Fig.~\ref{fig:sed_blazar} that Model 2 fails in describing the observed high-energy data. 
In fact, only by neglecting the inner gamma-gamma absorption (thin dot-dashed line) one could be able to attempt a successful fit of the data. The high compactness of this emission region implies a high opacity due to pair-production, and thus an unsuccessful fit. Furthermore, the synchrotron self-absorption affects the synchrotron component already in the UV range, thereby preventing a fit also in the optical and radio bands.

In Model 3, we try to explain the LAT data not as SSC, but rather as the high-energy tail of the synchrotron component. With respect to Model 1, we only modify the parameters of the electron distribution. The electrons are now parametrized by a broken-power-law function, $\gamma_{break}$ reproducing the peak, and $\gamma_{max}$ increased to reach the GeV band. 
In contrast to Model 2, Model 3 is not affected by internal absorption. It is characterized by an SSC component practically identical to Model 1 while also providing a good explanation of the GeV and sub-GeV data with the high-energy tail of the SYN spectrum. However, in order to fit the LAT data, Model 3 requires that the maximum Lorenz factor for the electrons is equal to $2\times10^9$. 
Such a large value exceeds by at least three orders of magnitude the standard maximum Lorentz factor of BL Lacs. 
Here, the maximum electron energy is constrained by the equality of the acceleration timescale and the fastest of the cooling timescales. At such high energies, the fastest cooling process is synchrotron radiation, which for electrons with $\gamma = 2\times10^9$ in a magnetic field $B = 0.001$ G is equal to $3.9\times10^5$ seconds. 
In order to see persistent synchrotron radiation from electrons at these energies, we need an acceleration process faster than that. The standard expression for the acceleration timescale in shocks is $t_{acc} = \psi^{-1} {m c\ \gamma}/{eB}$, where $\psi$ translates the efficiency of the acceleration process, and in blazars it is typically much smaller than unity \citep[see e.g.][]{InoueTakahara}. For $\gamma = 2\times10^9$, and $B = 0.001$ G, $t_{acc} = \psi^{-1} 1.1\times10^5$ seconds, i.e. shorter than the cooling timescale only assuming the highest possible efficiency $\psi = 1$. 
Any more realistic value of $\psi$ (0.1 and smaller), implies an impossible value of $\gamma$: electrons at these energies radiate faster than they can possibly be accelerated. 
Finally, by looking at Fig.~\ref{fig:sed_blazar}, one can also see that the SED loses the characteristic double-humped shape typical of the blazar population.

We conclude that, in the context of a one-zone leptonic model typically adopted to describe HBL objects \citep{Cerruti15}, only nonphysical parameterizations of 1E 1207.9+3945 can provide a sizable contribution to SRC-1 spectrum below 10 GeV and that, at most, the blazar contamination can be present at the highest energies, showing an SED typical of an HBL with a synchrotron peak in the X-rays and an SSC peak in the tens of GeV or more. 

Even though the blazar model we adopted generally provides good descriptions of HBL SEDs, we point out that leptonic models variations including additional physical ingredients, such as different acceleration mechanisms or complex source structures, might lead to unusual spectral features. 
In addition, the inclusion of hadrons might also allow us to provide a qualitatively better fit of the gamma-ray flux with the blazar.
Indeed, the possibility that the gamma-ray flux observed from the direction of NGC 4151 has a lepto-hadronic origin from the blazar 1E 1207.9+3945 was recently considered by \cite{Omeliukh2025}. However, the power in protons required to explain the gamma-ray flux at GeV via proton synchrotron is super-Eddington (the luminosities in \cite{Omeliukh2025} are provided in the reference frame of the emitting region), and results in a spectral energy distribution with three bumps that has never been observed in any blazar in any activity state. 
The results of the lepto-hadronic modeling, together with the physical parameters required to achieve a satisfactory fit, provide additional evidence that 1E 1207.9+3945 is unlikely to represent the dominant GeV gamma-ray source within SRC-1.

\section{Modeling the UFO in NGC~4151}\label{Sec: Model}

Seyfert 1 galaxies host an active core where the absorption column and/or the dusty torus alignment with the line of sight allows for the optical broad-emission lines to be directly observed, and for the X-rays to be also directly observable from the corona surrounding the SMBH. NGC~4151 is one the first Seyfert galaxies ever discovered~\citep{Seyfert43} and it features intermediate spectral properties between a Seyfert 1 and Seyfert 2~\citep{Osterbrock-1.5}. 
In addition, NGC~4151 hosts an X-ray UFO which has been previously detected and investigated in detail.

In order to constrain the observable properties of the UFO hosted in NGC~4151 we consider two sets of observations performed respectively by \textit{XMM-Newton}~\citep[][]{Tombesi_2010,Tombesi_2011,Tombesi2012} and \textit{Suzaku} X-ray satellites~\citep[][]{Gofford2013, Gofford-2015}, summarized in Table~\ref{table:parameters-from-obs}. 
The X-ray luminosity in the 2-10 keV band, $L_X$, and the wind speed, $u$, show good agreement among the two measurements. 
X-ray observations allow upper/lower limits to be placed on the UFO fast wind location given the measured parameters of the absorbing gas, albeit with large uncertainties. 
The maximum radius, $R_{\rm obs}^{\rm max}$, up to which high velocity gas is detected shows a wide range of values from a hundred gravitational radii up to parsec scale, while for the minimum radius, $R_{\rm obs}^{\rm min}$, consistent measurements are obtained at around a few tens of gravitational radii.
Similarly, the measured min (max) values for the mass loss rate, $\Dot{M}_{\rm min}$ ($\Dot{M}_{\rm max}$), and kinetic power, $\Dot{E}^{\rm min}_{\rm kin}$ ($\Dot{E}^{\rm max}_{\rm kin}$), show large uncertainties. 

Even if affected by relatively large uncertainties in its location, mass-loss rate and power, the UFO physical parameters allow us to identify the range of values to be considered for the following model.

\subsection{Cosmic ray acceleration and transport model}\label{subs: model}

The UFO is considered as a spherically symmetric and energy-conserving outflow featuring a bubble structure. 
The system is characterized as follows: 1) a wind shock (located at $R_{\rm sh}$) separates the innermost fast cool wind of constant velocity from the hot shocked wind, where the wind speed decreases as $\sim r^{-2}$; 2) a contact discontinuity (located at $R_{\rm cd}$) separates the hot shocked wind from the shocked ambient medium; 3) an outer forward shock (located at $R_{\rm fs}$) bounds the system. 

The main macroscopic parameters of an UFO are the mass loss rate, $\Dot{M}$, and the terminal wind speed $u$. The evolution, size and density profile of a system with given $\Dot{M}$ and $u$ are set by fixing the age $t_{\rm age}$ and the external medium density $n_{0}$ ~\citep[][]{Weaver77,Koo-McKee,Koo-McKee2,Faucher-Giguere-2012}. 
Following P23, the upstream magnetic field pressure is inferred as a small fraction $\epsilon_B =0.1$ of the ram pressure, while the downstream one is the result of compression of the perpendicular components of the upstream field at the wind termination shock. The magnetic field is also assumed to be characterized by a coherence length, $l_c =0.01 \, \rm pc$. The AGN photon field is computed following \cite{Marconi_2004_SED} with an X-ray luminosity within the range inferred from observations. 
The photon number density is assumed to dilute as the square of the distance from the SMBH. 
We also account for the infrared thermal component produced by a dusty torus \citep[][]{Mullaney11_torus}. 

\begin{table}
\caption{Parameters of the UFO in NGC\,4151 as inferred from X-ray observations with \textit{XMM-Newton}$^\dagger$ \cite{Tombesi_2010,Tombesi_2011,Tombesi2012} or \textit{Suzaku}$^*$ \cite{Gofford2013,Gofford-2015}.}
\vspace{0.5cm}
\label{table:parameters-from-obs}
\centering             
\renewcommand{\arraystretch}{1.05}
\linespread{1.05}\selectfont
\begin{tabular}{c|c|c|c}
\hline 
\makebox[1.6cm][c]{Parameter} &  \makebox[1.6cm][c]{\textit{XMM-Newton}${}^\dagger$} & \makebox[1.6cm][c]{\textit{Suzaku}${}^*$} &
\makebox[1.6cm][c]{Unit} \\
\hline
$L_X $ & $3.0$ & $2.0$ & $10^{42}\rm erg \, s^{-1}$   \\
$u $ &  $0.106$ & $0.055$ & c \\
$R_{\rm obs}^{\rm min}\,/\,R_{\rm obs}^{\rm max} $ & $10^{-4}\,/\,4 \cdot 10^{-4}$ & $2 \cdot 10^{-4}\,/\,0.4$ & $ \rm pc$  \\
$ \Dot{M}^{\rm min}\,/\,\Dot{M}^{\rm max}$  & $0.0025\,/\,0.04$ & $0.0006\,/\,0.25$ & $\rm M_{\odot} \, yr^{-1}$ \\
$ \Dot{E}^{\rm min}_{\rm kin} \,/\, \Dot{E}^{\rm max}_{\rm kin}$  & $0.8\,/\,12.6$ & $0.05\,/\,25.1$ & $10^{42} \, \rm erg \, s^{-1}$  \\
\hline 
\end{tabular}
\end{table}

DSA takes place at the wind termination shock, $R_{\rm sh}$, where a high Mach number can be found, while particles freely escape once they reach the forward shock. As in P23 we solve the stationary space-dependent transport equation in the whole system. The solution at $R_{\rm sh}$ can be expressed in the following compact form:
%
\begin{equation}
    f_{\rm sh}(p) = \mathcal{C} \left( \frac{p}{p_{\rm inj}} \right)^{-s} \exp{\left[-\Gamma_{\rm cut}(p)\right]},
\end{equation}
%
where $s$ is the slope of the injected particles -- assumed here as a free parameter -- $\mathcal{C}$ is a normalization constant set by $\xi_{\rm CR} \simeq 0.1$ (ratio between the CR pressure and the ram pressure at the wind shock) and $\Gamma_{\rm cut}$ is the cut-off function depending on the size of the system and the transport properties therein.
CR interactions with gas (pp) and radiation (p$\gamma$) produce hadronic emission of gamma rays and neutrinos \citep[our computational method is based on][]{Kelner_pp,Kelner_pg}.
We refer the interested reader to P23 for additional details on the acceleration and transport model.

\subsection{Application to NGC~4151}\label{subs: results}

In what follows we specialize our calculations to the case of NGC\,4151. In agreement with the observations, we consider an UFO of velocity $u \simeq 0.1c$ with a mass loss rate $\dot M \simeq 2\cdot10^{-2} M_\odot{\rm yr}^{-1}$. 
We explore a typical range of external density, $ 10^3 \lesssim n_0/{\rm cm^{-3}} \lesssim 10^6$ (see P23), and we derive the associated $t_{\rm age}$ and slope of the accelerated particles ($s$) requiring the predicted gamma rays to match the observations. Interestingly, the estimates of the emission size inferred from the X-rays provide a constraint on the available parameter space.
We find that a spectral index in momentum space $s=4.4$ is favored in order to properly fit the gamma-ray flux. 
We notice that, different from P23 and \cite{UFO-Fermi-LAT+Caprioli}, the soft gamma-ray spectrum observed requires an injection slope softer than the standard prediction of DSA. 
This can be motivated by a drift of the scattering centers in the downstream region of the shock~\citep{Caprioli_postcursor}. 
Alternative scenarios where the proton injection slope is harder, say $s=4$, could be a possible alternative if a leptonic cascade were contaminating the GeV range resulting in a softening of the observed spectrum. 
We leave a detailed UFO leptonic modeling to future investigations.

In general, we obtain that different combinations of $n_0$ and $t_{\rm age}$ can well explain the gamma-ray flux. Despite the parameter degeneracy, we identify three main configurations of the UFO that are compatible with the observations: A) sub-parsec-sized; B) parsec-sized; C) multi-parsec-sized.
Table~\ref{table:parameters-results} reports the specific values of the parameters relative to each configuration with the associated maximum energy that in all cases is $ \gtrsim 10^2 \, \rm PeV$.
We observe that scenarios with $n_0 \gg 10^5 \, \rm cm^{-3}$ are disfavored since they would result in a time variability of the UFO shorter than the Fermi-LAT observational time (see also Figure~\ref{fig:Fermi-LAT lightcurve}) and in a size of the order of $10^{-2} \, \rm pc$. Such a size is in tension with the $R^{\rm max}_{\rm obs}$ inferred from observations with Suzaku.

\begin{table}
\caption{Best fit parameters for the possible realizations of the UFO in NGC~4151.}
\vspace{0.5cm}
\label{table:parameters-results}
\centering             
\begin{tabular}{c|c|c|c|c}
\hline 
Model & $n_0 \, \rm [cm^{-3}]$ &  $t_{\rm age} \, \rm [yr]$ & $R_{\rm sh}$ -- $R_{\rm fs} \, \rm [pc]$ & {$E_{\rm max} \, \rm [EeV]$} \\
\hline 
  A & $10^{5}$ & 3.0 $\cdot 10^2$ &  0.1 -- 0.5 & 0.163  \\
  B & $10^{4}$ & 2.5 $\cdot 10^3$ &  0.6 -- 3.0 & 0.368  \\
  C & $10^{3}$ & 1.4 $\cdot 10^4$ &  2.5 -- 13 & 0.275  \\
 \hline
\end{tabular}
\end{table}

Figure~\ref{fig:Image_1} shows the gamma-ray and per-flavor neutrino spectra of our UFO model A as solid blue and dotted red lines, respectively. 
Model B and C do not differ from model A in the energy range where data are present. 
The emission is dominated by pp interactions in the shocked ambient medium layer (see also \cite{Mou_2015} for similar results) and follows the injection spectrum of the CR population. The gamma-ray spectrum is attenuated by gamma-gamma interactions \citep[][]{Aharonian_book2004} in the AGN photon field. 
As can be seen in the figure, our UFO model reproduces the inferred gamma-ray SED well within the 1\,$\sigma$ uncertainty band.
It is interesting to explore wether the predicted neutrino emission of the UFO could be tested by neutrino observatories, such as IceCube~\citep{IceCube:2022der}. 
IceCube's sensitivity for time-integrated muon-neutrino emission from a source located at declination $\delta\simeq 39^\circ$ is $E^2 F_{\nu_\mu+\bar\nu_\mu}(E) \simeq (4-8)\cdot10^{-13}\,{\rm TeV}\,{\rm cm}^{-2}\,{\rm s}^{-1}$ at a reference energy of 1\,TeV for spectral indices $\gamma=2.0-3.2$. 
As shown in Fig.~\ref{fig:Image_1}, at around $1$\,TeV the predicted per-flavor neutrino flux from the UFO is at the level of $E^2 F_{\nu_{\mu}+\Bar{\nu}_{\mu}}\simeq 5 \cdot 10^{-15}\,{\rm TeV}\,{\rm cm}^{-2}\,{\rm s}^{-1}$, {\it i.e.}~two orders of magnitude below IceCube's sensitivity.

\section{Discussion}
\label{Sec: Multim-Perspectives}

NGC~4151 is a unique laboratory to probe wind launching and acceleration processes in different energy regimes. 
In fact, its central engine is known to host a potential multi-phase outflow that produces a series of variable absorption features from the X-rays (2 -- 10 keV; \cite{sch02, Tombesi_2010, Gofford2013}) to the near infrared \citep[He {\scriptsize I};][]{wil16}. 
The investigation of such processes in this source, as already done in other nearby AGN such as Mrk\,231 \citep{fer15}, may thus contribute to shed additional light on the nature of multi-phase AGN outflows, that is still largely unconstrained so far \citep{cic18}. 
Together with the study of gamma rays, the search for neutrino emission from NGC~4151 constitutes a new powerful tool for a complementary investigation.

While the UFO model of NGC\,4151 can successfully account for the observed gamma-ray spectrum, it predicts neutrino emission that falls far below the sensitivity of current neutrino observatories. 
However, there exist alternative hadronic models of AGN that would allow to boost the neutrino emission compared to the relatively low gamma-ray flux. 
These models feature CR interactions in a region that is highly opaque to GeV gamma rays, such as the AGN nearest neighborhood~\citep[see e.g.][]{Inoue_2019,Inoue2020,Murase_AGN_cores,Foteini_NGC1068,Murase2022,Fiorillo2023,Fiorillo24_2,Padovani24,Lemoine2024} or a shocked environment by a combination of a successful and failed line driven winds close to the accretion disk \citep{Susumu_Inoue_NGC1068}. 
In particular, these models have been invoked to explain the multimessenger spectra of the Seyfert 2 galaxy NGC\,1068 where IceCube finds compelling evidence for neutrino emission~\citep{IceCube:2022der}.
Interestingly, NGC\,4151 is located at a distance of only $0.18^\circ$ from the fourth most significant hot spot in the search for northern neutrino point sources with spectral index $\alpha_{\nu} = 2.5$ in the same IceCube analysis (see supplement of \citep{IceCube:2022der} and \cite{Neronov_icecube,Abbasi24,IceCube-Seyfert}). 
In what follows, we estimate possible upper limits in the context of DSA for the neutrino emission in the AGN nearest neighborhood of NGC\,4151 considering the most optimistic scenario provided by calorimetric conditions.

We assume that a non-thermal population of protons in approximate equipartition with the X-rays is injected in the AGN vicinity. 
We also assume that such protons possess a total luminosity $L_{\rm CR}$ as large as a fraction $\eta_{\rm CR} \lesssim 3 \%$ of the AGN bolometric luminosity $L_{\rm bol}$. 
Following \cite{Inoue_2019} we assume that protons are injected with a spectral slope as hard as $\sim E^{-2}$ and a maximum energy ${E}'_{\rm max} \approx 1 \, \rm PeV$, where the maximum energy is in agreement with the assumed X-ray luminosity.
Adopting the notation of \cite{Murase-Guetta-Ahlers}, we can estimate an upper limit on the per-flavor neutrino flux in the calorimetric limit as:
%
\begin{equation}
  E^2 F_{\nu_\alpha} \lesssim \frac{K}{4(1+K)} \frac{\chi\eta_{\rm CR}L_{\rm bol}}{4 \pi D_{\rm L}^2}\,,
\end{equation}
%
where $\chi \equiv 1/\ln({E}_{\rm max}'/{\rm GeV})$ is a normalization factor assuming $E^{-2}$ CR spectra between $1 \, \rm GeV$ and ${E}_{\rm max}'$ and $K\approx 2(1)$ for pp (p$\gamma$) interactions. 
Assuming pp interactions, this gives an upper limit for NGC\,4151 of: 
%
\begin{equation}
  E^2 F_{\nu_\alpha}\lesssim 5\cdot10^{-13} \left(\frac{\eta_{\rm CR}}{0.03} \right) \left( \frac{\chi}{0.05} \right) \left( \frac{L_{\rm bol}}{10^{44} {\rm erg \, s^{-1}}} \right) \frac{\rm TeV}{\rm cm^{2} \, {\rm s}}\,. 
    \label{eq: calorim}
\end{equation}
%
This order of magnitude estimate illustrates that, the innermost region of the core of NGC~4151 can already produce a neutrino flux at the level of IceCube's present sensitivity level~\citep{IceCube:2022der}. 
Finally, we point out that stochastic acceleration~\citep{Murase_AGN_cores,Fiorillo24_2} and magnetic reconnection~\citep{Drury_2012,Kheirandish_2021,Fiorillo2023} could be alternative acceleration mechanisms in such environment. In particular, different from DSA, they could allow for injection spectra as hard as $\sim E^{-1}$. 
In such a scenario, the upper limit set in Equation~\eqref{eq: calorim} could be partially relaxed and, in the same parametric configuration, the neutrino flux could be peaking at TeV while being up to one order of magnitude larger in normalization at the peak, thereby saturating the flux that has been observed in \cite{Neronov_icecube} and \cite{Abbasi24,IceCube-Seyfert}.

{In \cite{Inoue_jet_and_corona_2023}, it is hypothesized that the slow ($v\lesssim 0.04 c$) radio jet observed in the core of NGC~4151 can also be a source of gamma rays. 
The radio-inferred jet power is of the order of $10^{42} \, \rm erg \, s^{-1}$ \citep{Williams1017}, so that $10-50\%$ of such a power would be necessary to be converted into relativistic electrons in order to reach the observed level of gamma rays. 
This would require an acceleration efficiency of electrons larger than the typical $1-10\%$ observed in the context of non-relativistic diffusive shock acceleration for both electrons and protons, where the latter typically dominate. In this context, the UFO interpretation appears more energetically plausible.} 

\section{Conclusions}
\label{Sec: Conclusions}

In this work we discuss the identification of gamma-ray emission from NGC\,4151, a nearby (15.8 Mpc) Seyfert 1.5 galaxy. 
The significance of the detection is 5.5~$\sigma$ and the association to NGC~4151 is supported by the spatial coincidence of the gamma-ray excess with the position of NGC~4151 and by the slope of the associated spectral energy distribution.
{NGC~4151 is located less than 5 arcmin from the blazar 1E~1217.93495. A careful spatial analysis leads us to the conclusion that the two objects cannot be distinguished by Fermi-LAT. 
On the other hand, a detailed spectral modeling of both sources indicates that NGC~4151 shall be considered the most physically-motivated counterpart of the measured gamma-ray flux, at least in the energy range below 10 GeV. 
In fact, only nonphysical {parametric configurations in the context of one-zone leptonic} blazar modelings can dominate the observed gamma-ray flux in such energy range. 
However, we do not discard the possibility of 1E~1217.93495 contaminating the source above 10 GeV.}

NGC~4151 hosts a mildly-relativistic UFO in its core region detected by Suzaku and XMM-Newton. 
Thus, we interpret the gamma-ray emission in the context of particle acceleration at the wind termination shock of the UFO where gamma-rays as well as HE neutrinos are produced via inelastic pp interactions in the shocked ambient medium. We notice, in particular, that the particle acceleration model is well constrained by the parameters obtained from X-ray observations and it is energetically motivated by standard parametric assumptions typical for the test-particle regime, namely the pressure of accelerated particles is $\lesssim 10 \%$ of the ram pressure at the shock. We find that particles could reach energies as high as $10^2 \,{\rm PeV}$ in this UFO.
We additionally highlight that from the star-formation rate (SFR) inferred for NGC~4151 ($0.25 \, \rm M_{\odot} \, yr^{-1}$ \cite{Erroz-Ferrer_2015}) one can predict the following gamma-ray luminosity $L_{\gamma}^{\rm (SFR)} \approx 2.9 \cdot 10^{38} \, \rm erg \, s^{-1}$ adopting the SFR-to-gamma correlation typical of star forming galaxies \citep[][]{Kornecki-1,Kornecki-2}. 
Interestingly, such luminosity is more than two orders of magnitude smaller than what we obtain with our measurement: $L_{\gamma} \approx 3.65 \cdot 10^{40} \, \rm erg \, s^{-1}$. This is a strong and clean indication that the gamma-ray emission is related to the AGN activity taking place in NGC~4151.
The HE neutrino flux associated to the gamma-ray observation is self-consistently computed in the context of the UFO model and we observe that it is about two orders of magnitude below the sensitivity of the IceCube observatory.

We note, however, that recent IceCube analyses 
find indications for neutrino emission in the vicinity of NGC\,4151 \cite{IceCube:2022der,Neronov_icecube,Abbasi24,IceCube-Seyfert}.
If NGC\,4151 was confirmed as a neutrino source 
at the level of the observations
it would imply that NGC\,4151 hosts a partially hidden accelerator with relatively low GeV counterpart, {\it e.g.}~close to the accretion disk with gamma-ray absorption in the AGN nearest neighborhood. 
We provide an estimate of the maximum neutrino emission achievable under calorimetric conditions and $E^{-2}$ spectral shape, and show that it is consistent with a flux as high as the present IceCube sensitivity level and the order of magnitude found in the observations.

While improving our analysis \cite{Inoue_jet_and_corona_2023} adopted our public preliminary results to propose the jet and the AGN corona as alternative gamma-ray emission sites {(see \S~\ref{Sec: Multim-Perspectives} for further comments). 
Interestingly, they also computed a neutrino flux produced in the corona of NGC~4151 and their result is consistent and very close with our upper limits possibly suggesting a calorimetric environment. 
{Depending on the size, the AGN corona could possibly contaminate the flux only around 100 MeV.}
In this context, there could be also another possible contamination to the gamma ray flux coming from a larger scale molecular outflow \citep[see e.g.][]{Lamastra_2016,Kraemer_outflow_2020}, while the star formation, as discussed, is unlikely to contribute.

NGC~4151 is an extremely interesting multi-messenger source and further observational campaigns are of crucial relevance in order to decipher its complex nature. 
Multi-wavelength observations could help understanding its constituents and characterize the non-thermal emission regions.
In particular, the TeV domain can provide timely and insightful information on the nature of the multi-messenger cosmic accelerator caught on act.

\acknowledgments

EP and MA acknowledge support by Villum Fonden (No.~18994). EP was also supported by the European Union’s Horizon 2020 research and innovation program under the Marie Sklodowska-Curie grant agreement No. 847523 ‘INTERACTIONS’. GP and EP were supported by Agence Nationale de la Recherche (grant ANR-21-CE31-0028). FGS was supported by the PRIN MIUR project ``ASTRI/CTA Data Challenge'' (PI: P. Caraveo), contract 298/2017. AL and FGS  acknowledge support from  INAF Mini-Grant  2023 “Particle Acceleration and multi-messenger emission in Non JEtted AGN”. EP is grateful to S. Bianchi and P. Padovani, for insightful discussions. The authors are grateful to F. Longo, G. Principe and L. Tibaldo for useful suggestions and comments on the analysis.

\bibliographystyle{JHEP}
\bibliography{references.bib} 

\end{document}